\documentclass{emulateapj}
\usepackage{epsf}
\usepackage{graphicx}
\usepackage{placeins}

\shorttitle{Radio Synchrotron Background Workshop Summary}
\shortauthors{Singal et al.}
\slugcomment{{\it PASP}, 2018, 130, 985}

\begin{document}
\title{THE RADIO SYNCHROTRON BACKGROUND:  CONFERENCE SUMMARY AND REPORT}

\author{J. Singal\altaffilmark{1}, J. Haider\altaffilmark{1}, M. Ajello\altaffilmark{2}, D. R. Ballantyne\altaffilmark{3}, E. Bunn\altaffilmark{1}, J. Condon\altaffilmark{4}, J. Dowell\altaffilmark{5}, D. Fixsen\altaffilmark{6},\\ N. Fornengo\altaffilmark{7}, B. Harms\altaffilmark{8}, G. Holder\altaffilmark{9}, E. Jones\altaffilmark{1}, K. Kellermann\altaffilmark{4}, A. Kogut\altaffilmark{6}, T. Linden\altaffilmark{10}, R. Monsalve\altaffilmark{11,12,13}, \\ P. Mertsch\altaffilmark{14}, E. Murphy\altaffilmark{4}, E. Orlando\altaffilmark{15}, M. Regis\altaffilmark{7}, D. Scott\altaffilmark{16}, T. Vernstrom\altaffilmark{17}, L. Xu\altaffilmark{16}}

\altaffiltext{1}{Department of Physics, University of Richmond\\28 Westhampton Way, Richmond, VA 23173, USA}
\altaffiltext{2}{Department of Physics \& Astronomy, Clemson University, 118 Kinard Laboratory, Clemson, SC 29634-0978, USA}
\altaffiltext{3}{School of Physics, Georgia Institute of Technology, Atlanta, GA 30332-0430, USA}
\altaffiltext{4}{National Radio Astronomy Observatory, 520 Edgemont Road, Charlottesville, VA 22903-2475, USA}
\altaffiltext{5}{Department of Physics \& Astronomy, University of New Mexico, 1919 Lomas Blvd. NE, Albuquerque, NM 87131-0001, USA}
\altaffiltext{6}{NASA Goddard Space Flight Center, 8800 Greenbelt Rd, Greenbelt, MD 20771, USA}
\altaffiltext{7}{Department of Theoretical Physics, University of Torino and INFN, Via Giuria 1, 10125 Turin, Italy}
\altaffiltext{8}{Department of Physics \& Astronomy, University of Alabama, 514 University Blvd.
Tuscaloosa, AL 35487-0324, USA}
\altaffiltext{9}{Department of Physics, University of Illinois at Urbana-Champaign, 1110 West Green Street, Urbana, IL 61801-3003, USA}
\altaffiltext{10}{Department of Physics, The Ohio State University, 191 West Woodruff Avenue, Columbus, OH 43210, USA}
\altaffiltext{11}{Center for Astrophysics and Space Astronomy, University of Colorado at Boulder, 389 UCB, Boulder, Colorado 80309-0389, USA}
\altaffiltext{12}{School of Earth and Space Exploration, Arizona State University, Tempe, AZ 85287, USA}
\altaffiltext{13}{Facultad de Ingenier{\'i}a, Universidad Cat{\'o}lica de la Santísima Concepci{\'o}n, Alonso de Ribera 2850, Concepci{\'o}n, Chile}
\altaffiltext{14}{Theoretical Particle Physics and Cosmology, Niels Bohr Institute, Blegdamsvej 17, 2100 Copenhagen \O, Denmark}
\altaffiltext{15}{Department of Physics, Stanford University, 382 Via Pueblo Mall, Stanford, CA 94305-4060, USA}
\altaffiltext{16}{Department of Physics \& Astronomy, University of British Columbia, 6224 Agricultural Road, Vancouver, BC V6T 1Z1, Canada}
\altaffiltext{17}{Dunlap Institute for Astronomy \& Astrophysics, University of Toronto, 50 St. George Street, Toronto, ON M5S 3H4, Canada }

\email{jsingal@richmond.edu}

\begin{abstract}
We summarize the radio synchrotron background workshop that took place July 19--21, 2017 at the University of Richmond.  This first scientific meeting dedicated to the topic was convened because current measurements of the diffuse radio monopole reveal a surface brightness that is several times higher than can be straightforwardly explained by known Galactic and extragalactic sources and processes, rendering it by far the least well understood photon background at present.   It was the conclusion of a majority of the participants that the radio monopole level is at or near that reported by the ARCADE~2 experiment and inferred from several absolutely calibrated zero level lower frequency radio measurements, and unanimously agreed that the production of this level of surface brightness, if confirmed, represents a major outstanding question in astrophysics.  The workshop reached a consensus on the next priorities for investigations of the radio synchrotron background.  
\end{abstract}


\section{Introduction} \label{intro}
The radio synchrotron background is an astrophysical phenomenon that has been a subject of interest to many in the community in recent years.  The measured surface brightness level of the radio monopole is several times that implied by the source counts of discrete extragalactic radio sources, which are the explanations for the observed monopoles at infrared, optical/UV, X-ray, and gamma-ray wavelengths.  In addition, various possible Galactic and extragalactic production mechanisms are highly constrained.

Although a bright high Galactic latitude diffuse radio monopole had been reported as early as the 1960s \citep{Costain60,Bridle67} and was seen in data from the 1980s \citep{Phillips81, Beuermann85}, early estimates of the high-latitude sky simply assumed that the observed intensity was some mixture of an extragalactic background from radio point sources with the remainder allocated to a Galactic contribution, and neither of these was particularly well constrained at the time.  Renewed interest came with the publication of results from the ARCADE~2 experiment \citep{Fixsen11,Kogut11,Singal11}.  Combining the ARCADE~2 balloon-based absolute spectrum data from 3--90 GHz with absolutely calibrated zero level single-dish degree-scale resolution radio surveys at lower frequencies \citep[e.g][]{Haslam} re-confirmed a bright radio synchrotron monopole level.  This level is several times brighter than had widely been predicted, as estimates of the Galactic monopole level and combined extagalactic point sources have become more firmly constrained in recent decades \citep[e.g.][]{Vernstrom11}.  

These results, although consistent between ARCADE~2 in the low GHz range and lower frequency absolutely calibrated zero level radio surveys in the MHz range, are not without controversy.  ARCADE~2 was designed as an absolute measurement but had limited sky coverage, while the lower frequency radio surveys were not designed with such an absolute zero-level calibration as a primary goal, and did not tend to stress the intricacies of their zero-level calibration when reporting results.  Still, the agreement between ARCADE 2 and absolutely calibrated zero level MHz radio surveys necessitates investigation, because such a result, if true, would potentially imply very interesting consequences for our understanding of Galactic and/or extragalactic radio emission.

\begin{figure}
\includegraphics[width=3.5in]{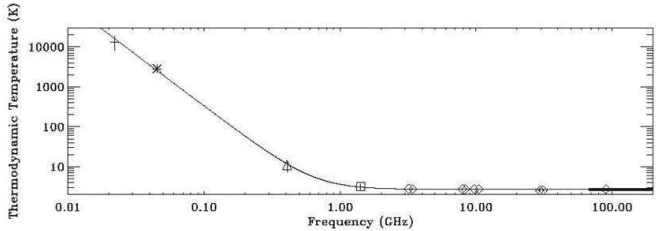}
\caption{Spectrum of the sky monopole in radiometric temperature units, reproduced from \citet{Seiffert11}, as measured by several different instruments or surveys.  The spectrum shows a clear power-law rise at frequencies below $\sim$10 GHz, above the otherwise dominant cosmic microwave background (CMB) level.   Diamonds are measurements from ARCADE~2 \citep{Fixsen11}, the square point is the measurement from \citet{RR86}, the triangle point is the measurement of \citet{Haslam}, the star point is the measurement of \citet{Maeda99}, and the plus point is the measurement of \citet{Roger99}.  Higher frequency measurements from {\it COBE}-FIRAS \citep{FM02} are represented as the thick solid line.  } 
\label{CRB}
\end{figure}

\begin{figure}
\includegraphics[width=3.5in]{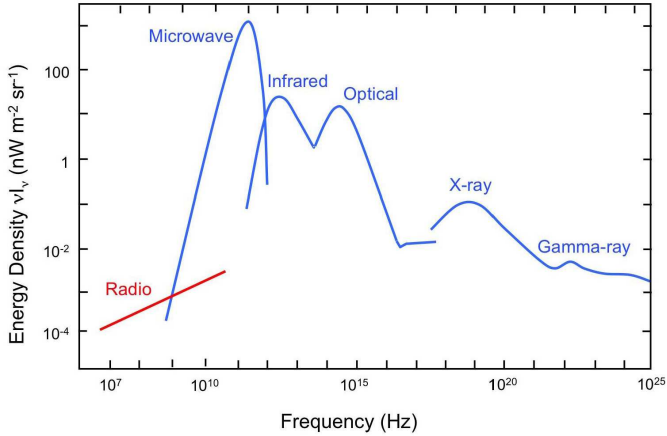}
\caption{Spectral energy surface brightness density of the photon backgrounds in the universe, with the level of the radio synchrotron background taken from ARCADE~2 and absolutely calibrated zero level radio maps as reported by \citet{Fixsen11}.} 
\label{CRB2}
\end{figure}

The radio synchrotron monopole level as reported by ARCADE~2 and lower frequency surveys is spatially uniform enough to be considered a ``background,'' thus it would join the astrophysical backgrounds known in all other regions of the electromagnetic spectrum. As there are many reasons to believe the diffuse radio monopole is extragalactic and in fact cosmological, many have taken to referring to it as the ``cosmic radio background.'' As we strive for neutrality on this question this work will refer to it as the radio synchrotron background. The reported radio synchrotron monopole spectrum as a function of frequency in radiometric temperature units is shown in Figure \ref{CRB}, while a comparison with the photon backgrounds in other wavebands is shown in spectral energy surface brightness units in Figure \ref{CRB2}.

If it is indeed at the level determined from ARCADE~2 and lower frequency absolutely calibrated zero level radio surveys as reported by \citet{Fixsen11}, the origin of the radio synchrotron background would be one of the mysteries of contemporary astrophysics. It is difficult to produce the observed level of surface brightness by known processes without violating existing constraints. A brief review of some recent literature on the subject follows:

Some authors \citep[e.g.][]{SC13} have proposed that the background originates from a radio halo surrounding the Milky Way Galaxy.  Such a large halo has been seen as an outcome in detailed studies of cosmic ray propagation models \citep[e.g.][]{OS13} and is possible in scenarios where the Galaxy has a significant wind \citep[e.g.][]{Breitschwerdt02,Biermann10}.  There are important difficulties with a Galactic origin, however.  With the $\sim$1\,$\mu$G magnetic fields determined by radio source rotation measures to be present in the Galactic halo, the same electrons energetic enough to produce the radio synchrotron background at the observed level would also overproduce the observed X-ray background through inverse-Compton emission \citep{Singal10}.  Also the observed correlation between radio and singly ionized carbon line (\ion{C}{2}) emission \citep{Kogut11} would also imply an overproduction of the observed level of emission from that line above observed levels.  Furthermore, independent detailed modeling of the structure of the diffuse radio emission at different frequencies does not support such a large halo \citep{Fornengo14}.  Lastly, although galaxies with large radio emitting halos do exist \citep[e.g.][]{Weigert17}, a halo of the necessary size and emissivity would make our Galaxy anomalous among nearby similar spiral galaxies \citep{RB1}.  

However an extragalactic origin for the radio synchrotron background also presents many challenges.  Several authors have considered deep radio source counts \citep{Vernstrom11,Vernstrom14,Condon12} and concluded that the background would require an incredibly numerous new population of radio sources far below the flux densities currently probed by deep interferometric surveys.  These results are in agreement with others who have probed whether active galactic nuclei \citep[AGN --- ][]{Ball2011} or other objects \citep{Singal10} are numerous enough.  Some authors have noted that the far-infrared background would be overproduced above observed levels if the radio synchrotron background were produced by sources that follow the known correlation between radio and far-infrared emission in galaxies \citep{Ponente11,YL12}, while others have observed that the correlation may evolve with redshift and have noted the implications for the radio background \citep{IvisonA,IvisonB,Magnelli15}. Other authors have investigated the anisotropy of the radio synchrotron background and shown that arcminute scale fluctuations are sufficiently small that if the sources trace the large scale structure of the Universe the background must be produced by objects at redshifts greater than $z\sim$5 \citep{Holder14}, and some of the simplest high redshift mechanisms have been ruled out \citep{CV14}.  

Such constraints have led authors to consider origins such as annihilating dark matter in halos or filaments \citep{Fornengo11,Hooper14,FL14} or ultracompact halos \citep{Yang13}, ``dark'' stars in the early universe \citep{Spolyar09}, supernovae of massive population~III stars \citep{Biermann14}, emission from Alf\'{e}n re-acceleration in merging galaxy clusters \citep{FL15}, an enhancement in the local bubble \citep{Sun08}, and dense nuggets of quarks \citep{LZ13}. Topics that inform and are informed by considerations of the radio synchrotron background include radio source counts, X-ray source counts, quasar luminosity functions, cosmic ray propagation, Galactic and extragalactic magnetic fields, the radio to far-infrared correlation, dark matter annihilation, polarization of foregrounds in microwave background maps, population~III stars, and cluster mergers.

A workshop on radio synchrotron background was merited given the level of interest in exploring its origin and that it touches on so many contemporary issues in astrophysics. This report presents a summary of the major discussions and conclusions of the workshop for the rest of the scientific community. In \S \ref{details} we give an overview of the logistical details of the meeting. \S \ref{summary} gives a summary of individual presentations in the workshop, and some important overall conclusions from the various discussions are presented in \S \ref{conc}.

\section{Meeting Details}\label{details}
This workshop was sponsored in part by a grant from the National Science Foundation and with support from the University of Richmond College of Arts \& Sciences and the University of Richmond Physics department. The scientific organizing committee was responsible for all technical, planning, and logistical details, and was comprised of the chair Jack Singal (University of Richmond), Ted Bunn (University of Richmond), Alan Kogut (NASA Goddard Space Flight Center), and Ken Kellermann (National Radio Astronomy Observatory). Individuals who participated in the workshop are listed in Table \ref{participants}.

\begin{deluxetable}{ll}
\tablecaption{Participants\label{participants}} 

\tablehead{\colhead{Name} & \colhead{Affiliation} \\ 
\colhead{} & \colhead{} } 

\startdata
David Ballantyne & Georgia Tech \\
Ted Bunn & University of Richmond \\
Jim Condon & National Radio Astronomy Observatory \\
Jayce Dowell & University of New Mexico \\
Dale Fixsen & NASA Goddard Space Flight Center \\
Nicolao Fornengo & University of Turin \\
Jibran Haider & University of Richmond \\
Ben Harms & University of Alabama \\
Gil Holder & University of Illinois at Urbana-Champaign \\
Evan Jones & University of Richmond \\
Ken Kellermann & National Radio Astronomy Observatory \\
Alan Kogut & NASA Goddard Space Flight Center \\
Tim Linden & The Ohio State University \\
Raul Monsalve & University of Colorado \\
Philipp Mertsch & Neils Bohr Institute \\
Eric Murphy & National Radio Astronomy Observatory \\
Elena Orlando & Stanford University \\
Marco Regis & University of Turin \\
Douglas Scott & University of British Columbia \\
Jack Singal & University of Richmond \\
Tessa Vernstrom & University of Toronto \\
Lenon Xu & University of British Columbia \\
\enddata
\end{deluxetable}
\FloatBarrier

The workshop website\footnote{http://physics.richmond.edu/rsb-conference} contains a repository of the program and all presentation slides.  

\section{Summary of Individual Presentations}\label{summary}

\subsection{Introductory Talk - Challenges and Opportunities --- Alan Kogut}
\label{Kogut1}

\begin{figure}
\includegraphics[width=3.5in]{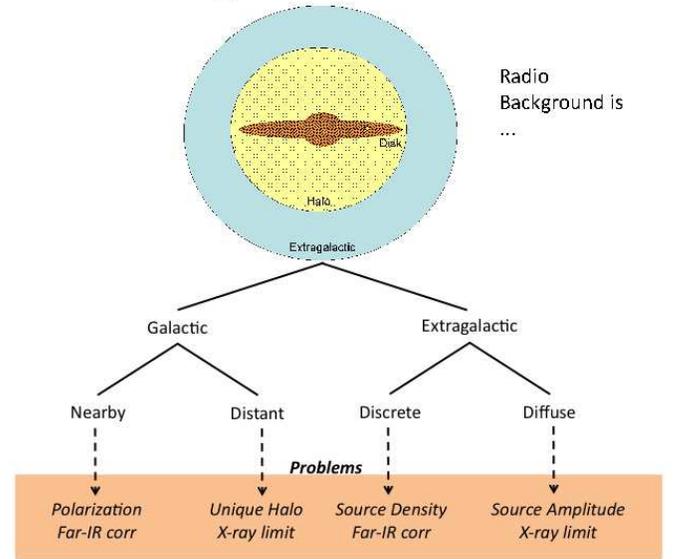}
\caption{Schematic of possible origins for the radio synchrotron background and their issues.  } 
\label{flowchart}
\end{figure}

This talk first summarized the measurements from ARCADE~2 and the low frequency absolutely calibrated zero level sky maps, noting the presence of a large monopole level seen in radio map data as early as the 1960s \citep{Costain60,Bridle67}.  A brief introduction to the possible origin scenarios and a few of their pitfalls is visualized in Figure \ref{flowchart}.  Some of these potential problems were discussed in detail in subsequent presentations, including, for the Galactic halo origin, the presentation by Jack Singal (\S \ref{Singal}), for discrete extragalactic sources the presentations by Jim Condon (\S \ref{Condon}) and Eric Murphy (\S \ref{Murphy}), and for diffuse extragalactic sources the presentations by David Ballantyne (\S \ref{Ballantyne}) and Tessa Vernstrom (\S \ref{Vernstrom}).   Additionally, the possibility of an origin in the local bubble region of the Galaxy seems to be constrained by the lack of a large-scale polarization signal seen in GHz-frequency maps from e.g. WMAP.  The observed radio sky is strikingly depolarized, which is difficult to explain if the bulk of the emission comes from local (Galactic) sources.

Some options for future measurements to validate the level and spectrum of the radio synchrotron background observed by ARCADE~2 and the single dish radio surveys include the following.

(i) An absolutely calibrated zero level large-sky-coverage map at $\sim$120\,MHz, where the background sky temperature is roughly 300\,K, equal to the ground emission temperature, so that ground pick-up would be less of a challenge

(ii) An absolutely calibrated zero level large-sky-coverage map at $\sim$300\,MHz, which could be achieved with a custom feed horn mounted on the existing Green Bank telescope

(iii) A non-absolutely calibrated large-sky-coverage map at one of the ARCADE~2 observing frequencies around 3 or 8\,GHz, where the spatially overlapping ARCADE~2 data could be used to provide a zero-point calibration

(iv) Polarization measurements at frequencies above 5\,GHz to further determine the structure of the Galactic magnetic field, with an eye toward constraining a local origin that should have some degree of large-scale polarization.  These could include Faraday rotation measurements and/or small-scale synchrotron polarization measurements in the faintest parts of the sky.

\subsection{ARCADE and Other Measurements --- Dale Fixsen}\label{Fixsen}

The level of the radio synchrotron background is established via the ARCADE 2 balloon-borne absolute radiometer, and calibrated single-dish radio surveys.  The ARCADE~2 measurement provides a good zero-point calibration \citep[e.g.][]{Singal11}.   (1) the measurement was made from a balloon which eliminates most of the air and almost all of the water vapor above; (2) there are no windows and the entire instrument is kept at ~1.5\,K, minimizing stray input from the instrument and its surroundings; and (3) perhaps most importantly, the measurement of the sky is compared directly to a cold load at the same temperature as the sky. Thus ARCADE~2 provided a measurement good to a few mK over a small patch of sky, including the Galactic plane and some high latitudes, at 3, 8, 10, 30 and 90\,GHz.

These measurements of the radio monopole are consistent with previous measurements by \citet{Roger99} at 22\,MHz, \citet{Maeda99} at 45\,MHz, \citet{Haslam} at 408\,MHz, and \citet{RR86} at 1420\,MHz, with a spectral index of $-$2.6.  It is important to note that with these single dish radio surveys, one cannot simply take the coldest pixel as the monopole temperature, since instrumental noise creates a distribution of pixel values around the coldest patch of the sky --- rather one must create a histogram of the coldest pixels and fit a Gaussian with the mean value indicating the actual monopole temperature.  ARCADE~2 also agrees with {\it COBE}-FIRAS measurements \citep[e.g.][]{FM02} in its higher frequency bands.  These measurements are well described by a cosmic microwave background (CMB) temperature of 2.725\,K along with a spatially varying Galactic synchrotron component with a mean spectral index of $-$2.6 and a smooth, nearly uniform component. The spatially-varying Galactic component distribution is similar to the infrared emission, or the singly ionized carbon (\ion{C}{2}) emission.  This is similar to measurements of other spiral galaxies.

The smooth componet also has a synchrotron spectrum with an index of $-$2.6. All three components appear with any combination of data.  Unfortunately the ARCADE~2 data cover only a small patch of sky, and the low frequency data zero point is poorly determined and described.  The {\it COBE}-FIRAS data have a good absolute calibration but do not extend to low enough frequency to really pull out the synchrotron emission. The uniform component is a factor of five or six times what is expected from known point sources (see e.g. \S \ref{Condon}).

It would be fruitful to make measurements over the range 20\,MHz to 3\,GHz with well calibrated zero points and gains either over large patches of the sky or small high latitude patches. Measurements in the 100--120\,MHz range are particularly interesting because the sky temperature in this range is aound 300\,K making the job of matching the sky relatively simple.  It also means that spillover onto the ground is less of a problem.

\subsection{The LWA1 Sky Survey --- Jayce Dowell}\label{Dowell}

To get a better understanding of the nature of the radio synchrotron background one needs to have reliable measurements at a variety of frequencies.  In particular, low frequency measurements, below 200\,MHz, are important in determining the spectral index of the background and for searching for evidence of a break in the power law or a turn over. However, there are few existing surveys in this frequency range and most of the surveys that do exist have poorly documented calibration procedures and unknown zero point offsets.  Thus, more modern surveys that have a demonstrable absolute calibration are key to understanding the origin of this background.

One such survey is the LWA1 Low Frequency Sky Survey \citep[LLFSS ---][]{Dowell17}.  The first station of the Long Wavelength Array, LWA1, is a dipole array consisting of 256 dual polarization antennas spread over a 110$\times$100\,m area.  LWA1 also has five outrigger dipoles that provide baselines up to 500\,m for calibration \citep{Tay12}.  The LWA1 supports beam forming, and two all-dipole raw voltage data capture modes, one narrow band and continuous (Transient Buffer Narrowband; TBN) and one wide band but with a 0.03\% duty cycle (Transient Buffer Wideband; TBW).  For the survey, data between 10 and 88\,MHz were captured using the TBW mode from the central 256 antennas in 61\,ms bursts roughly every 15 minutes over two days.  The resulting snapshots were correlated and imaged using the LWA Software Library \citep{Dowell12}.  The LLFSS covers the radio sky north of $-$40$^{\circ}$ declination in nine frequency bands between 35 and 80\,MHz.  Across this range the survey has a spatial resolution of  $\sim$5$^{\circ}$ to 2$^{\circ}$, respectively.  The total power calibration of the survey is provided by the precision radiometers developed for the Large aperture Experiment to detect the Dark Ages \citep[LEDA ---][]{Price09}.  These radiometers are the outrigger antennas and have modified front-end electronics and a temperature-controlled noise source which are designed to provide an antenna temperature that is calibrated to better than 1\,K.  In addition, the survey approach of combining multiple snapshots of the visible hemisphere of the sky allows for the estimation of the two-dimensional temperature uncertainties across the sky.

Using this survey, along with existing surveys and new results from total power (sky-averaged) measurements from cosmic dawn and epoch of reionization experiments (see e.g. \S \ref{Monsalve}), it should be possible to make precise measurements of the radio background level below 200\,MHz.  These measurements will also be of use in aiding our understanding of the Galaxy and may help to disentangle whether the radio synchrotron background is of primarily Galactic or extragalactic origin.

\subsection{The Contribution of Galaxies to the Background at 3\,GHz --- Jim Condon}\label{Condon}

In spite of pushing to deeper flux density limits, radio source count measurements have given a consistent picture of the extragalactic source population for decades.  Most radio sources are extragalactic, and divide into two populations, namely AGN-drven sources such as quasars and radio galaxies, and star-forming galaxies, characterized by different flux densities at which their normalized number counts peak.  These two classes comprised the extragalactic objects observed in the 3rd Cambridge Revised Catalogue in the 1960s \citep[e.g.][]{Bennett62}, although they were not distinguished at the time.  As radio source count surveys have probed deeper flux densities, the objects tallied have gone from being primarily those at higher luminosities to those at lower luminosities, but the two class division has remained, with AGN-driven sources peaking in their integrated 1.4\,GHz contribution to the total radio surface brightness of the sky at around $\sim$0.1\,Jy; and star-forming galaxies becoming dominant below $\sim$1\,mJy \citep[e.g.][]{Condon84}.   To the limits currently probed, these radio sources add up to approximately one-fifth of the radio synchrotron background level reported by \citet{Fixsen11} and discussed in \S \ref{intro}, with two-thirds of that integrated amount coming from AGN-driven sources and one-third coming from star-forming galaxies.

At the deepest flux densities currently probed, star-forming galaxies dominate the radio sources.  The source counts power-law slope of these objects at these deep flux densities is an essential input to estimating their integrated contribution to the total radio surface brightness via extrapolation.  Recent deep surveys have obtained quite discrepant results on this question, with the survey of \citet{OM08} showing a much larger number of sources at deep flux densities than some other recent surveys \citep[e.g.][]{MC85}.  

With the goals of (1) determining an accurate source count down to $S\sim 1\,\mu$Jy, (2) resolving the radio sky background contributed by galaxies, and (3) constraining possible source populations that could produce the high level of the radio synchrotron background, \citet{Condon12} used the Karl G. Jansky Very Large Array (VLA) to make a very sensitive (RMS noise $\sigma_{\rm n} \approx 1\,\mu$Jy beam$^{-1}$) low-resolution (8'' full-width-at-half-maximum) confusion-limited sky image covering one primary beam area at S band (2 - 4 GHz) that is centered on the same Lockman Hole field mapped by  \citet{OM08}.  At this resolution Milky Way-sized sources at cosmological distances are unresolved and the level of source confusion is comparable to the RMS noise.   This level of source confusion lends itself to $P(D)$ analysis \citep[e.g.][]{Scheuer57}, where the known RMS noise and measured image variance are used to determine the underlying confused source distribution, since the measured variance is a quadratic sum of the other two.  This technique can constrain the source counts at flux desnties well below those where sources are directly detected.  With such an analysis, \citet{Condon12} determined that the source counts at flux densities below 10\,$\mu$Jy are constrained to be well below that determined by \citet{OM08}.  Because of their narrow beam and high resolution, \citet{OM08} probably overstated the source size correction of the weakest sources which resulted in increasing their determined object counts at the lowest flux density levels.  

\begin{figure}
\includegraphics[width=3.5in]{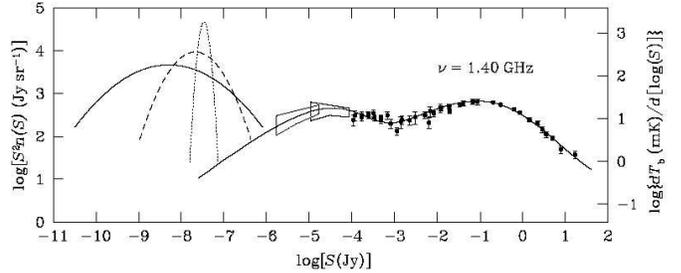}
\caption{From \citet{Condon12} and discussed in \S \ref{Condon}.  The points with error bars show the measured source counts from several surveys summarized in that work.  The solid line running largely through the data points is a model of the source counts resulting from a combination of AGN-driven sources and star-forming galaxies presented in \citet{Condon84}.  The boxes indicate the allowed range of source counts consistent with the $P(D)$ analyses in \citet{MC85} (higher fluxes) and \citet{Condon12}.  The three parabolas at nanoJy levels correspond to models for a hypothetical new population of radio sources with Gaussian distributions in logarithmic flux density (with different standard deviations) able to produce the reported radio synchrotron background brightness temperature of \citet{Fixsen11}.  Even in the limit of a very narrow flux density distribution, the number of sources per square arcmin is $N>6 \times 10^4$ arcmin$^{-2}$, which greatly exceeds the $N\sim$10$^3$ arcmin$^{-2}$ galaxies brighter than magnitude +29 in the Hubble Ultra Deep Field.  } 
\label{Cfig}
\end{figure}

The extrapolation of source counts to lower flux densities given the constraints from the $P(D)$ analysis in \citet{Condon12} is consistent with the standard two-population source model and a falling integrated contribution to the total radio sky surface brightness with decreasing source flux density.  This conclusion is also in agreement with the low-flux radio source counts that would be inferred from applying the radio-far-infrared correlation to deep infrared source counts.  It seems from radio source count considerations that neither AGN-driven sources or star-forming galaxies can provide the bulk of the observed level of the radio synchrotron background seen by e.g. \citet{Fixsen11}.  \citet{Condon12} also quantified the flux range and number density of hypothetical populations that could provide the missing surface brightness, given the constraints from the $P(D)$ analysis, and these are shown Figure \ref{Cfig}.  It is clear that such a population would have to be either very numerous or very sharply peaked in its flux distribution, which are characteristics of no currently known population.

\subsection{The Contribution of AGN to the Excess Cosmic Radio Background at 1.4 GHz --- David Ballantyne}\label{Ballantyne}

Radio-loud or jetted AGN dominate the AGN-driven radio source number counts at 1.4\,GHz flux densities greater than $\sim$1\,mJy. However, the majority of non-jetted or radio-quiet AGN also produce radio flux and these begin to dominate the AGN-driven source counts below around 0.1\,mJy. In addition to their core emission, star formation within the host galaxy of radio-quiet AGN will also contribute to their radio flux \citep{Ball09}. Therefore, when computing the radio background brightness temperature due to AGN-driven sources, it is important to include the contribution from both radio-quiet AGN and potential host-galaxy star formation.

\citet{Ball2011} performed this experiment using the method of \citet{Ball09} that predicted the 1.4 GHz radio number count from both radio-loud and radio-quiet AGN (plus host star formation). The method relies on using the X-ray luminosity function to track the evolution of the total luminous AGN population, including both radio-loud and radio-quiet AGN. Phenomological conversions between X-ray luminosity and radio luminosity were used to calculate the radio flux densities from the model AGN population. 

During this workshop an updated calculation of the AGN contribution to the 1.4\,GHz radio background was presented.  The latest results presented here are found in \citet{Draper17}.  This calculation made use of a more recent X-ray luminosity function \citep{Ueda14}, and conversion from X-ray to radio luminosity \citep{Pan15}. The radio luminosity for a specific star-formation rate is given by the relationship described by \citet{Murphy11}. The core emission of the radio-quiet AGN is assumed to have a photon index of $\alpha=0.2$, while star formation and radio-loud AGN are assumed to have radio emission with a photon index of $\alpha=0.7$. The luminosity-dependent radio-loud fraction can be constrained using the latest AGN number counts from \citet{Padovani15}. The \citet{Smolcic17} results were also included to compare against the models. The brightness temperature was computed by integrating the counts from 100 nJy to 10 mJy. The temperature above 10 mJy (51 mK) is taken from Figure 13 of \citet{Condon12} and is added to this result to obtain the final 1.4 GHz brightness temperature due to AGN.

The model that includes only AGN with no host star formation gives a total $T_{\rm b}=67$\,mK. This is slightly higher than the \citet{Condon12} result; however, this model significantly underpredicts the radio-quiet AGN number counts. Removing this deficit by adding host galaxy star-formation (that includes both a moderate amount of $z$-dependence and AGN luminosity dependence) yields the best fit model and $T_{\rm b}=76$\,mK. A simple star-formation model where all AGN have a host-galaxy star-formation rate of $2.7$~${\rm M}_{\odot}\,yr^{-1}$ produces a number count model that just barely brushes the upper-bounds of the data and is considered the ``maximal'' model. The temperature of this model is $T_{\rm b}=88$\,mK. Taking this last value as an upper-limit and the ARCADE~2 measurement of the 1.4\,GHz background temperature of 480\,mK \citep{Fixsen11} suggests that the AGN contribution to the 1.4\,GHz background is $< 18$\%. Following \citet{Ball2011} and adding up the known contributions to the background (see their Table 2) means that known classes of sources can only explain up to a maximum of $\sim67$\% of the background at 1.4\,GHz. 

\subsection{Diffuse Extragalactic Syncrotron Emission and the Radio Background --- Tessa Vernstrom}\label{Vernstrom}

The extragalactic portion of the radio background can be looked at in terms of two components: (1) emission from discrete sources, or individual galaxies, such as AGN cores and star-forming galaxies; and (2) emission from more diffuse sources. These diffuse sources can be individual galaxy halos, extended emission from AGN (such as jets or lobes), halos and relics in galaxy clusters and groups, or emission from filaments (i.e. the synchrotron cosmic web).  At 1.4\,GHz there are approximately 100 diffuse cluster sources known \citep[giant and mini radio halos and radio relics,][]{Feretti12}, which amounts to about $0.1\,$mK in background temperature. The sources that have been directly detected are generally brighter (flux density $> 1\,$mJy) and found in higher mass clusters. These sources are difficult to detect, since they tend to be faint, have low surface brightness, and be spread over large angular scales (which can be problematic for radio interferometers to measure); additionally there is contamination from bright foreground sources (the Galaxy or bright point sources) and confusion with faint point sources. These issues present real challenges for measuring diffuse emission from lower mass clusters or groups and filaments. 

\begin{figure}
\includegraphics[width=3.5in]{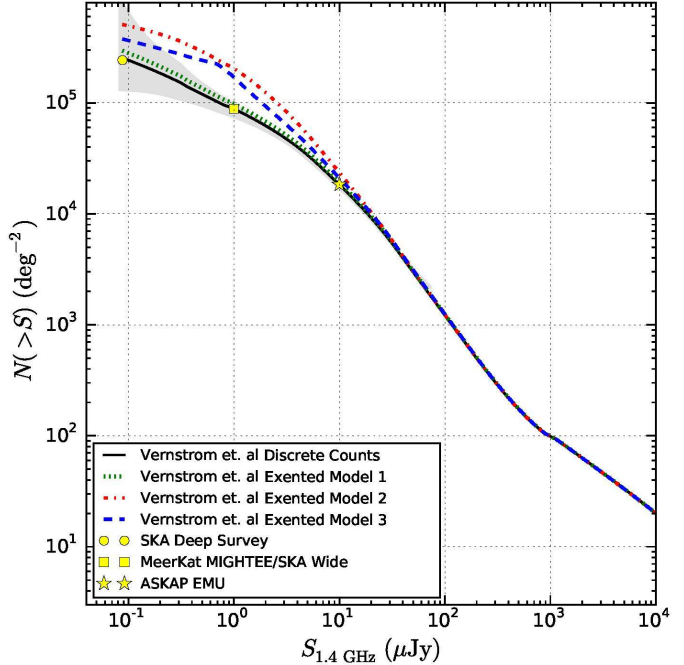}
\caption{Models for inetgrated source counts at $1.4\,$GHz discussed in \S \ref{Vernstrom}. The solid black line shows the integrated discrete source count model from \citet{Vernstrom14}. The colored lines show the discrete model added to the extended source count models from \citet{Vernstrom15}. The yellow points show the predicted survey depths for the SKA and SKA pathfinders MeerKat and ASKAP.  } 
\label{diffusesources}
\end{figure}

While source confusion is usually a problem, \citet{Vernstrom15} were able to use the $P(D)$ technique to estimate upper limits on the amount of diffuse emission present in the ELIAS S1 field. They found a range of background temperatures from diffuse emission at 1.75\,GHz  of 5--15\,mK, depending on the assumed model.  These implications for deep radio source counts of these estimates are shown in Figure \ref{diffusesources}.  They also showed that the technique can be applied to constraining source count models of cluster sources or other types of sources (e.g synchrotron emission in galactic halos coming from dark matter annihilation). 

\citet{Vernstrom17} and \cite{Brown17} used cross-correlation analysis to obtain upper limits on the synchrotron cosmic web emission at 180\,MHz and 2.3\,GHz, respectively. These limits correspond to approximately 1\,mK at 1.4\,GHz. While the cross-correlation method does enhance a signal buried in the noise, there can still be foreground contamination from the Galaxy and point sources, and the results are highly model dependent.

These works have found a 1.4\,GHz upper limit to the extragalactic background temperature from diffuse radio sources at $1.4\,$GHz of $\le 15\,$mK, or about 3\% of the radio synchrotron background level reported in \citet{Fixsen11}. Measurements of the X-ray background place limits on the radio temperature from diffuse emission (due to the synchrotron - inverse Compton connection) of about 5\% of the \citet{Fixsen11} temperature. Future surveys with new data may soon yield tighter constraints, howeve, the use and development of statistical methods of measurement are key (due to foreground issues such as confusion), as well as the development of new models to interpret the statistical results.

\subsection{Dark Matter, Alfv{\'e}n Acceleration, and the ARCADE~2 Excess --- Tim Linden}\label{Linden}

Common models of weakly interacting massive particle (WIMP) dark matter (DM) feature annihilation and/or decay channels into electrons and positrons at the GeV scale.  Radiation from these energetic annihilation products has been suggested as possible sources of the 511\,keV positron excess emission \citep{Knodlseder05}, GeV gamma-ray excess emission \citep[e.g.][]{GH09}, and microwave ``haze'' seen at the Galactic center \citep[e.g.][]{Finkbeiner04}, as well as the positron excess seen near Earth \citep{Adriani11}, and the radio synchrotron background \citep{Fornengo11,LZ13,Hooper14,FL14}.

The total integrated synchrotron contribution from DM annihilation (or decay) products is a function of the annihilation (or decay) rate, squared (or linear) DM density, and magnetic field strength.  It is straightforward to produce the observed level of the radio synchrotron background reported in \citet{Fixsen11} and discussed in \S \ref{intro} with properly parameterized DM models, with the synchrotron spectrum primarily governed by the DM particle mass and particular annihilation (or decay) channel, and absolute level governed primarily by the DM annihilation (or decay) rate, substructure model, and magnetic field strength and distribution.  However models that are sufficient to produce the observed level of the radio synchrotron background also tend to overproduce the observed level of the X-ray and gamma-ray backgrounds, via both direct gamma-ray production in annihilation channels and inverse-Compton upscattering of the CMB photons (as discussed in \S \ref{Singal}).   Furthermore, as discussed in \S \ref{Holder}, the smoothness of the diffuse radio emission indicates that the sources of the radio synchrotron background cannot trace the large scale structure at low redshifts, which the DM distribution does, unless the radio sources are larger than 2\,arcmin.  

Models in which DM annihilation or decay is a significant component of the radio synchrotron background therefore should ``go early'' (to the early Universe) or ``go big'' (to extended sources).  The former models tend to be somewhat fine tuned.  The latter is an intriguing option because the largest clusters have DM halos that are larger than 2\,arcmin.  Furthermore, when dark matter substructure is considered, there can be large boost factors for annihilation away from the cluster centers.  The major complication that needs to be addressed is the magnetic field strength in the annihilation regions, which needs to be sufficiently large to avoid an inverse-Compton overproduction of other observed backgrounds.

A possible solution would be a model where local magnetic field enhancements trace cluster dark matter substructure, even far away from cluster centers.  As investigated in \citet{FL14}, in such regions of locally enhanced magnetic field, the magnetic field turbulence can give rise to Alfv{\'e}n waves, which re-accelerate the electrons, leading to a locally enhanced synchtrotron emissivity.  Since the synchrotron emission would be happening only in these localized regions of the energetic electrons and enhanced magnetic fields, the overall inverse-Compton emission level for a given amount of synchrotron could be suppressed.  

\begin{figure}
\includegraphics[width=3.5in]{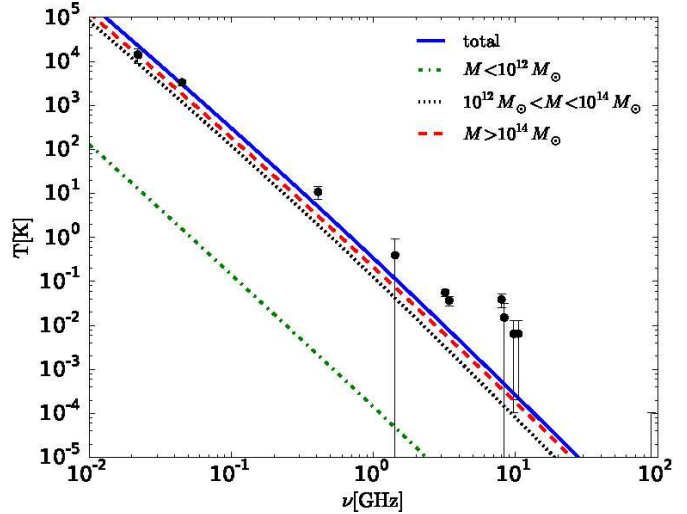}
\caption{From \citet{FL15}, showing the calculated level of synchrotron emission from cluster mergers with a population of electrons re-accelerated by Alfv{\'e}n waves in regions of locally high magnetic fields, given a particular turbulence spectrum and Reynolds number for the hydrodynamics of the cluster merger, as discussed in \S \ref{Linden}.  The contributions from clusters of different mass ranges is shown, as well as the observed level of the radio synchrotron background from \citet{Fixsen11}. A flatter spectrum can be obtained by hardening the electron injection in some energy range, such as having electron secondaries produced through hadronic interactions} 
\label{Lfig}
\end{figure}

This scenario also suggests the possibility of not needing the DM as the ultimate source of electrons.  Rather, ambient (non-thermal) electrons could theoretically be re-accelerated by Alfv{\'e}n waves in regions of locally high magnetic fields in clusters, giving rise to local regions of synchrotron emission.  Indeed, large-scale diffuse radio emission from clusters has been detected \citep{BR11}, and radio relics have been observed far from cluster centers with no inverse-Compton X-ray emission \citep{George15}.  Furthermore, cluster mergers would be a natural source of magnetic turbulence and shocks to initially accelerate electrons.  \citet{FL15} have calculated a resulting observed radio synchrotron absolute spectrum from such merger events, assuming a turbulence spectrum and Reynolds number for the hydrodynamics of the cluster merger.  The level can match that of the radio synchrotron background reported by \citet{Fixsen11}, although the spectrum is softer, as shown in Figure \ref{Lfig}.  A harder spectrum matching that of the radio synchrotron background can be obtained by hardening the electron injection in some energy range, such as having electron secondaries produced through hadronic interactions.  Additionally most of the brightness is generated by the most massive (and spatially largest) clusters, which means that the sources would be spatially extended enough to avoid the smoothness constraint (discussed in \S \ref{Holder}), and only 0.5--5\% of the total thermal power of the cluster need be in magnetic turbulence.  

There will be some significant halo-to-halo variation in the electron spectrum produced via the Alfv{\'e}n reacceleration mechanism. Interestingly, as shown in \citet{FL15} both the total synchrotron flux and the synchrotron emission spectrum are strongly dependent on the magnetic field strength, but the dependences are inverted. In particular, the maximum electron energy (which determines the synchrotron spectral hardness), is proportional to $B^{4/3}$, while the total Alfv{\'e}n power, which determines the total synchrotron luminosity is proportional to $B^{-4/3}$, $B$ being the magnetic field strength.  The eventual dependence of this synchrotron spectrum on the maximum electron energy is shown in Figure 7 of that work.  The physics behind this is that the conversion of turbulent energy into electron acceleration is more inefficient at high Alfv{\'e}n velocities.  Thus, the brightest individual clusters are expected to have relatively low magnetic fields and should not have highly energetic electrons, and would thus have a softer synchrotron spectrum than the average system, meaning that while some clusters may be observed to have a synchrotron spectrum that is steeper than that of the radio synchrotron background, if clusters as a whole contribute significantly to the background it is those with a softer spectrum that will contribute the most.  It should also be noted that in the above re-acceleration scenario the re-accelerated electrons cannot be those that emit the cluster X-ray emission through thermal Bremsstrahlung \citep[e.g.][]{PE08}.  However, there are multiple indications that galaxy clusters contain a significant non-thermal electron population \citep[e.g. from Iron line observations as explored in][]{Prokhorov13}. A number of different acceleration mechanisms, such as AGN, star formation, or collisional shocks during the merger event, may produce these electrons \citep{BJ14}. If their re-acceleration emission were indeed a significant component of the radio synchrotron background, the spectrum of these electrons would have to be relatively soft, and therefore their number density would be extremely high.  Thus only a small fraction of the $\sim$MeV non-thermal electron population would need to be accelerated to GeV energies in order to produce the observed signal.

\subsection{Backgrounds Due to the Formation of Supermassive Black Holes in the Early Universe --- Benjamin Harms}\label{Harms}

This talk presented a model \citep{Biermann14} which accounts for many of the observed massive particle and massless wave backgrounds. This model is based on the assumption that super-massive stars form in the early universe ($z\sim$50) when the metallicity is low. These super-massive population~III stars are treated as scaled-up versions of supernovae in that they explode, leaving super-massive black holes, while the remnants of the explosions are the primary source population of the isotropic radio frequency background.  

This model satisfies all known constraints on clustering, source number, and source flux levels, and is able to quantitatively account for the observed radio flux density, the observed number of sources per solid angle and the observed dependencies of the radio frequency spectrum on energy and frequency. As an example, the observed radio spectrum implies that the particle spectrum varies with energy as $E^{-2.2}$.  In this model the particle spectrum varies as $E^{-2.24}$. This model requires the far infrared emission to be negligible, matching the observed infrared spectrum.  A mechanism to evade excessive inverse-Compton upscattering of the ambient backgrounds whose energy density scale as $(1+z)^4$ (see discussion in \S \ref{Singal}) has also been proposed, namely that the inverse-Compton emission has a different time dependence than the radio synchrotron emission in the expansion of an individual population~III supernova remnant blast wave.

In addition to being able to account for the quantities associated with the radio frequency spectra, this model provides an explanation of the sources of ultra-high energy cosmic rays and neutrinos. The precessing jet which results from the merger of two super-massive black holes is the source of the observed UHECRs and neutrinos. A possible explanation of blazars, which are the probable sources of the observed high energy gamma rays, is that they are what is detected when the jet from such a merger is pointing directly toward Earth. This model also predicts the existence of ultra-low frequency gravitational waves, which are yet to be detected.

\subsection{Resolving the Extragalactic Gamma-ray Background --- Marco Ajello (Presented by Jack Singal)}\label{Ajello}

Understanding of the extragalactic gamma-ray background (EGRB) has progressed greatly in recent years, and can potentially provide lessons for investigating the radio synchrotron background.  The {\it Fermi} Gamma-ray Space Telescope's Large Area Telescope (LAT) is a unique instrument in being able to simultaneously measure the level of an extragalactic background and resolve it into sources \citep[e.g.][]{Ack16,Ackermann15}.  The level of the EGRB in the range of energies from 100\,GeV to 800\,GeV can apparently be entirely achieved by the contributions from blazars, star-forming galaxies, and radio galaxies.  

The contribution from any given source class can be estimated in two ways, either by utilizing source counts or utilizing luminosity functions.  The former has the advantage of being straightforwardly determined at observed fluxes but has an uncertain extrapolation to fluxes below those observed.  The latter has the possible advantage of utilizing additional relevant information --- the redshift evolution, but the disadvantage of a more complicated integration involving more variables and functions over those variables to get the total contribution to the observed surface brightness.  Recent works from the {\it Fermi}-LAT collaboration \citep[e.g.][]{Ajello15} have carried out these estimations with {\it Fermi}-LAT observations for blazars and found blazars in total produce around 50\% of the EGRB and produce the cutoff in the EGRB above $\sim$100\,GeV.  Flat spectrum radio quasar (FSRQ)-type blazars produce a larger share of this contribution than BL Lac-type blazars.  Other works have used different analysis techniques on the {\it Fermi}-LAT observations, with e.g. \citet{BP2} and \citet{BP3} in particular featuring rough agreement with the \citet{Ajello15} results although with significantly larger uncertainties.  

The {\it Fermi}-LAT has detected 15 radio galaxies and 8 star-forming galaxies in gamma rays \citep{Ackermann12,DiM14}.  The latter display a strong correlation between the gamma-ray and infrared emissions, and so the total star-forming galaxy contribution to the EGRB can be estimated, with the well-characterized infrared luminosity function of galaxies and a scaling factor \citep{Ackermann12}.  The observed radio galaxies display a correlation between the gamma-ray luminosity and the core radio luminosity at 5 GHz, so their total contribution to the EGRB can be estimated with the well-characterized radio luminosity function of radio galaxies and a scaling \citep{DiM14}.  The contributions of both star-forming galaxies and radio galaxies are found to be $\sim$25\% of the EGRB.  Thus the EGRB represents, in likely its entirety, the sum of blazars, star-forming galaxies, and radio galaxies.

\subsection{SPIDERz -- A Support Vector Machine for Photometric Redshift Estimation --- Evan Jones}\label{Jones}

Obtaining accurate and well understood estimates of millions of galaxies' redshifts from photometry in a limited number of wavelength bands --- so-called ``photometric redshift'' or ``photo-$z$s'' --- is a crucial challenge for current and especially upcoming large-scale astronomical surveys and their science goals such as cosmological probes like weak lensing \citep[e.g.][]{Hearin10} and determining the large-scale distribution of structure in the universe, which is relevant to several considerations related to the radio synchrotron background examined in this report.  This talk discussed the development of a custom support vector machine empirical machine learning algorithm for photo-z estimation called SPIDERz \citep[SuPport vector classification for IDEntifying Redshifts --- ][]{JS1}.  SPIDERz is useful for exploring various aspects of empirical photo-$z$ estimation because it naturally outputs an effective estimated distribution of redshift probability (EPDF) for each galaxy.  

SPIDERz has been tested with data sets with known redshifts and realistic photometry spanning wide redshift ranges and found that it performs competitively with popular photo-z codes in straightforward discrete photo-$z$ estimation.  A new study explores using SPIDERz's individual galaxy EPDFs to predict potential problematic galaxies that are likely to have poor photo-$z$ estimates, often called ``catastrophic outliers''  \citep{JS2}.  By applying essentially two criteria to the EPDFs (the ratio of the height of secondary peak(s) in the probability distribution to the primary peak and the distances of the secondary peak(s) from the primary peak in redshift) one can identify a large fraction of the catastrophic outliers while flagging only a small fraction of non-outliers.  Those interested in minimizing the effects of photo-z errors in analyses of large-scale surveys are encouraged to explore using this method of identifying potential catastrophic outliers, and to consider utilizing SPIDERz for photo-$z$ explorations.

\subsection{Waveband Luminosity Correlations in Multiwavelength Flux-Limited Data --- Jibran Haider}\label{Haider}

Determining the intrinsic correlations between the emissions in different wavelength bands (e.g. optical, radio, infrared, X-ray, etc.) for astrophysical sources is important for addressing a large variety of scientific questions, including the nature of the correlation between radio and far-infrared emission.  When dealing with multiwavelength observations of astrophysical sources the question often arises whether the emissions in different wavebands are truly intrinsically or only observationally correlated.  A common practice is to plot luminosities in two bands against each other and determine the correlation empirically.  This has been done for e.g. the radio far-infrared correlation.  However, for extragalactic sources with a range of redshifts coming from data which are flux-limited, the fact that lower (higher) luminosities in both bands are dominated by sources at lower (higher) redshifts introduces a significant correlation in the observed luminosities.  The situation is even more complicated, however, than merely issues with observational selection effects.  In addition to the selection effects, other factors, most prominently the similarities or differences in the redshift evolutions of the luminosity functions, can complicate the situation and induce correlations between different waveband luminosities that are not physically real.  As shown in \citet{galp} and \citet{corrpaper} using simulated data sets, luminosities that are completely intrinsically uncorrelated can appear very tightly correlated in flux-limited multiwavelength observations.

That forthcoming work explores the efficacy of the use of the Pearson correlation coefficient (PCC) and Pearson {\it partial} correlation coeficient (PPCC) between the luminosities in bins of redshift to access the true degree of correlation between different waveband emissions which are affected by the observational and redshift evolution effects discussed above.  This is done in part with simulated two-band population data sets with a variety of known local (i.e. prior to any redshift evolution) intrinsic luminosity correlations and redshift evolutions which were then subjected to realistic flux limited surveys in two bands to form simulated `observed' data sets on which the techniques could be tested.  It is found that by considering both the PCC and PPCC for both the raw and local luminosities, one can accurately deduce the presence and degree of intrinsic correlation between the luminosities.  When applied to a real flux limited optical-radio quasar data set \citep[discussed in ][]{QP2} and a real optical-mid-infrared quasar data set \citep[discussed in ][]{QP3}, this technique reveals that the optical and mid-infrared luminosities are very highly intrinsically correlated while the optical and radio data are only marginally intrinsically correlated.  While noting that this particular analysis can only reveal the degree of intrinsic correlation between the luminosities and not the nature of the best-fit correlation function itself, the very high degree of correlation seen in this analysis between mid-infrared and optical luminosities in quasars lends support to the picture of the dusty tori which often surround accretion disks being heated primarily by the emission from the accretion disks.  The significantly weaker correlation between radio and optical luminosities can be taken to support the notion that radio emission is affected by both the accretion disk size and the black hole spin, the latter perhaps being most important.  These results support an overall picture where black hole size determines accretion disk size and luminosity, which then dominates the optical emission and becomes the primary driver of infrared emission via heating of the torus, while both black hole spin and size determine jet strength and therefore the radio luminosity.  These results will be presented in a forthcoming paper \citep{corrpaper}.

\subsection{The Diffuse Galactic and Extragalactic Radio Emission --- Nicolao Fornengo}\label{Fornengo}

The radio synchrotron background is a superposition of the brightness of the Galactic diffuse monopole and the extragalactic component, and the relative size of these two is a subject of much debate.  \citet{Fornengo11} revisited this, employing state-of-the-art models for the Galactic foreground.  That work collected the most complete calibrated sky maps at radio frequencies ranging from 22\,MHz to 2326\,MHz, which all feature at least 50\% sky coverage and performed detailed theoretical modeling of the Galactic foreground emission. These models included the Galactic diffuse synchrotron radiation, the Galactic thermal bremsstrahlung (free-free) and single sources, plus an isotropic extragalactic background temperature $T_{\rm E}$ which was the goal of the analysis.

As the Galactic plane was masked in this analysis, the free-free component is not that crucial.  The synchrotron component is modeled with broken power-law cosmic-ray injection spectra with indexes $\beta_{\rm inj,nuc,i}$ and $\beta_{\rm inj,e,i}$ for nuclei and electrons, respectively, where $i$=1,2 for either the high or low energy power law index, and with the breaks at 9 GeV for nuclei and 4 GeV for electrons; a radial profile from the observed supernova remnant distribution and a vertical profile power law of the form exp($z/z_{\rm s}$) with $z_{\rm s}$=0.2\,kpc.  The cosmic ray propagation is modeled with GALPROP \citep[e.g.][]{SM98} with a magnetic field structure model from Jansson and Farrar \citep{JF1,JF2}.  The latter consists of 
(1) large scale regular field components
(a) from the disk,
(b) a toroidal halo component, and 
(c) an out-of-plane component;
(2) a striated random component, which is traced by synchrotron polarization or the lack thereof; and
(3) a small scale turbulent random component which is traced by synchrotron temperature.
To allow flexibility in the mid-to-high latitude emission, the random component magnitude is allowed to have a general form
\begin{equation}
B(R,z)=B_0 \, e^{{{-(R-R_{\rm T})} / {R_{\rm B}}}} \, e^{ {{-|z|} / {z_{\rm B}} }}
\end{equation}
with $R_{\rm T}$=8.5\,kpc, $R_{\rm B}$=30\,kpc, and $z_{\rm B}$ varying according to the model.  

Eight different models were run with a range of different $z_{\rm B}$, $\beta_{\rm inj,nuc,i}$, $\beta_{\rm inj,e,i}$, and $B_{\rm 0}$ values.  All models reproduce the observed cosmic ray energy spectra for electrons, antiprotons, and Boron/Carbon.  The radio maps from 22 to 2326\,MHz were then fit to these models of diffuse Galactic emission and an extragalactic monopole temperature $T_{\rm E}$.  Different methods were adopted to assess uncertainties associated with the Galactic modelling and with the treatment of the identified sources. The latter was performed in two ways: by adopting masks or by adding templates to the fit. 

\begin{figure}
\includegraphics[width=3.5in]{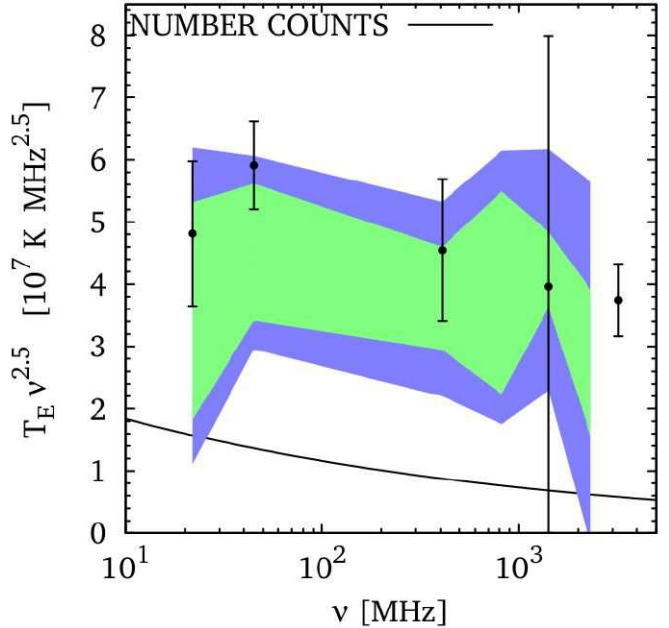}
\caption{Temperature versus frequency of the extragalactic component $T_{\rm E}$ calculated as discussed in \S \ref{Fornengo} with 1 and 2$\sigma$ propagated uncertainties arising from different model parameters indicated by the green and blue bands respectively. The points with error bars show the extragalactic temperature reported by \citet{Fixsen11}, while the lower curve shows the integrated temperature arising from the source counts calculated in \citet{Gervasi08}. } 
\label{TE}
\end{figure}

The results for $T_{\rm E}$ obtained in the different models were all above from what can be estimated from number counts of observed extragalactic sources, including extrapolation of those number counts to very low flux densities, and are in line with the extragalactic temperatures reported by \citet{Fixsen11}.  Figure \ref{TE} shows the reconstructed values of $T_{\rm E}$ at a range of frequencies for various model parameters along with the extragalactic temperatures reported by \citet{Fixsen11} and an estimate of the integrated emission from extragalactic source counts from \citet{Gervasi08}.  All results for $T_{\rm E}$ were compatible among the different methods, with a moderate scatter.  It is important to note that the reconstructed $T_{\rm E}$ exceeds the estimate of the integrated emission from extragalactic source counts by a factor larger than three for those maps with the largest fraction of sky available (45\,MHz, 408\,MHz, and 1.4\, GHz).  This analysis confirms the presence a high extragalactic radio synchrotron background level.  Simply put, observational maps of the diffuse radio emission point towards a puzzling extragalactic background.  The extragalactic interpretation is quite interesting, since its origin could be linked to dark matter-driven emission \citep{Fornengo11}.

\subsection{Synchrotron Propagation in the Context of Cosmic Ray Propagation Models --- Elena Orlando}\label{Orlando}

The Galactic radio synchrotron emission is composed mainly of diffuse emission produced by cosmic-ray (CR) electrons and positrons spiraling along the Galactic magnetic fields.  This emission, as detected by radio telescopes, is integrated along the line of sight, hence its level depends on the electrons and the magnetic fields at each position in the Galaxy.  The synchrotron intensity depends on the intensity of the magnetic field and the density of the electrons, while the synchrotron spectrum depends on the spectrum of the CR electrons.  Modeling this emission is very challenging, due to the large uncertainties in the electrons and magnetic fields throughout the Galaxy.  To complicate the modeling, while CRs propagate in the Galaxy their spectrum is modified by energy losses (e.g. synchrotron itself and inverse-Compton) and energy-dependent diffusion, and it can also be affected by re-acceleration processes and production of secondary CRs. These secondary CRs can produce significance synchrotron radiation at frequencies below a few hundred MHz \citep{O17}. Moreover their spectrum and distribution in the Galaxy is different from the primary electrons. Below particle energies of a few GeV, the local interstellar electron spectrum cannot be directly measured, because CRs are affected by solar modulation, so extrapolations of the locally measured spectrum to the spectrum throughout the Galaxy is very challenging. Free-free absorption is also a factor, especially in the Galactic plane.  All of these effects challenge our knowledge of the Galactic synchrotron emission, and make the usual synchrotron power-law approximation inadequate for the entire Galaxy and frequency range, calling for more sophisticated propagation models.

The GALPROP simulation package \citep[e.g.][]{SM98} models the propagation of cosmic rays through the Galaxy, with the goal of having a complete description of CRs and the associated diffuse emission from radio to gamma rays.  The ingredients for any particular simulation are an injected spectrum for CR species and propagation parameters, the CR source spatial distribution, the magnetic field structure, the gas distribution, and the interstellar radiation field for inverse-Compton losses.  

The level of the diffuse emission seen in low frequency radio surveys (those discussed in e.g. \S \ref{Fixsen}) between $|10^{\circ}|$ and $|40^{\circ}|$ Galactic latitude and longitudes away from the Galactic center ($40^{\circ} < l < 180^{\circ}$ and $180^{\circ} < l < 320^{\circ}$) can be fit with appropriate GALPROP parameter choices, including a break in the injected electron energy spectrum, and has provided important information about CRs and their propagation \citep{Strong11}.  

More recently the spatial distribution of the total and polarized synchrotron emission has been studied in the context of propagation models \citep{OS13}, where the GALPROP code has been extended to include polarization, absorption, and a 3D model of the magnetic field, with its regular, anisotropic, and random components. Different CR source distributions, propagation parameters, halo sizes, and magnetic field models were investigated, and were found to significantly change the spectral and spatial map of the calculated Galactic synchrotron emission.

That work showed that further tuning of the input parameters is required to also fit the polarization versus pixel longitude seen in WMAP data and the temperature vs. pixel longitude seen in the Haslam 408\,MHz map \citep{Haslam}.  These require the three-component magnetic field --- large-scale regular, striated random, and turbulent random --- that is also discussed in e.g. \S \ref{Fornengo}.  Moreover, a CR halo size larger than usually assumed ($\sim$10 kpc) was found to explain most of the monopole component in the 408\,MHz map and WMAP.  Although the interpretation may be affected by large uncertainty in the modeling, the analysis shows that most of the high Galactic latitude radio synchrotron monopole excess can have a Galactic origin.  The question remains as to whether the Milky Way indeed has an emissive halo of this size, which has been modeled by others as well \citep[e.g.][]{SC13}, and some constraints on which are mentioned in \S \ref{Singal}.  

\subsection{The Diffuse Radio Emission and Dark Matter --- Marco Regis}\label{Regis}

\begin{figure}
\includegraphics[width=0.35\textwidth]{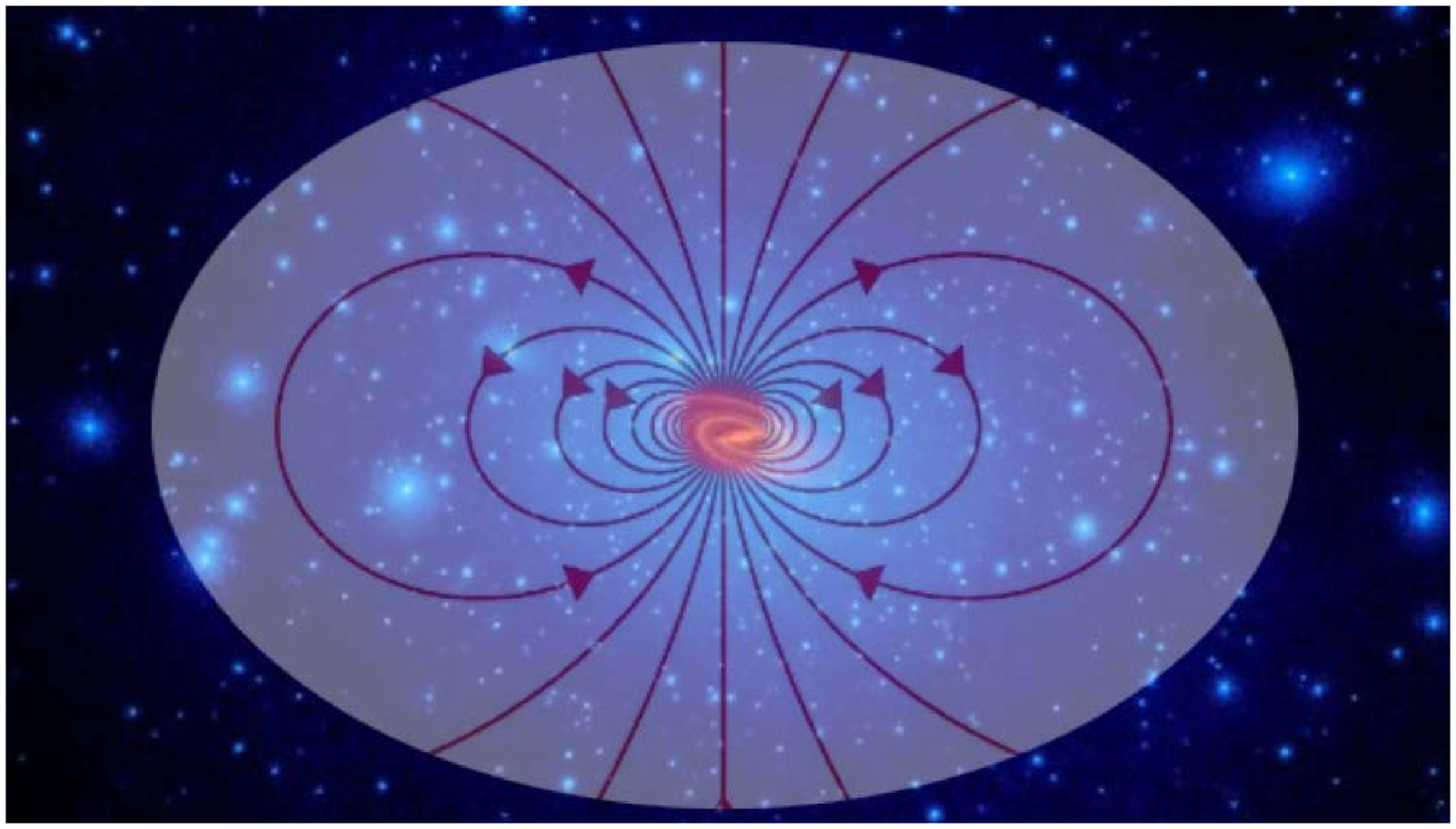}
\hspace{3mm}
\includegraphics[width=0.35\textwidth]{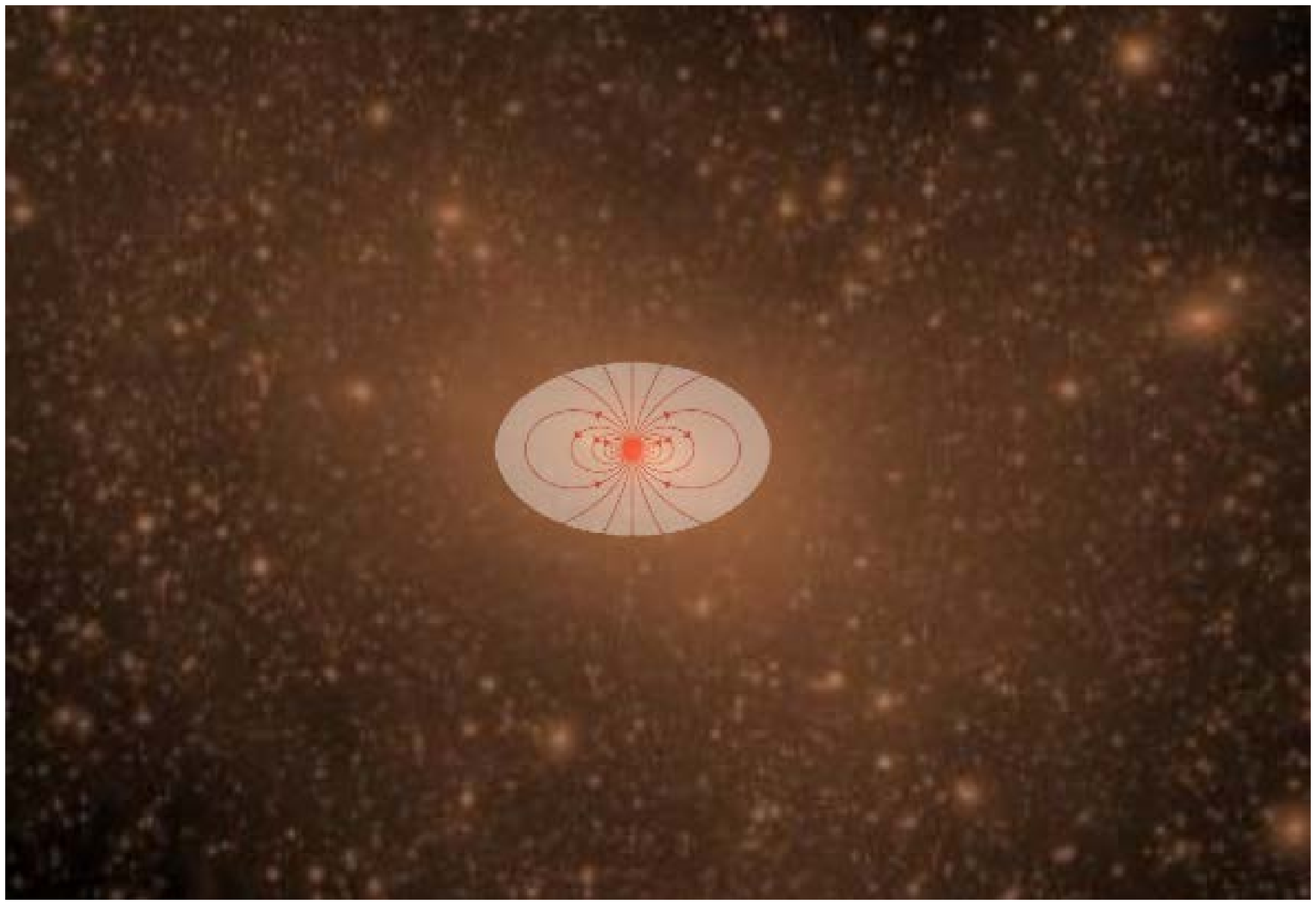}
\hspace{3mm}
\caption{Sketch of the two scenarios discussed in \S \ref{Regis} that can lead to a significant contribution from dark matter annihilation/decay to the radio synchrotron background.  The top panel depicts a galaxy with the magnetic field extending over the dark matter halo, while the bottom panel depicts a significant magnetic field present in a dark structure with no baryons.  Bright spots here correspond to dark matter over-densities.}
\label{fig:DMcartoon}
\end{figure}

If (and only if) the physical pictures sketched in Figure~\ref{fig:DMcartoon} are realized in nature, then the annihilation/decay of DM particles in extragalactic halos is generically predicted (in the framework of weakly interacting massive particles (WIMPs) as DM candidates) to induce synchrotron radiation at the level of the measured radio synchrotron background \citep{Fornengo11}. Indeed, the (weak) coupling of dark matter with Standard Model particles implies a significant injection of high-energy electrons and positrons (with details of their spectrum depending on the specific particle dark matter model). The question is whether or not these particles can generate a relevant synchrotron flux, and, in particular, if this can happen in regions where the baryonic emission is low or absent. This would require (a) magnetic fields extending much further than the baryonic matter distribution in galaxies (together with DM profiles having large cores); and/or (b) significant magnetic fields in dark structures, i.e. in the dark matter structures that do not have significant star formation today.

In order to test for these hypotheses, both single-target observations and statistical analysis of the radio synchrotron background can be applied.  Dwarf spheroidal galaxies (dSph) of the Local Group are among the most promising targets for particle dark matter searches. They are the most dark matter dominated structures known and have a large dark matter content.  Recent radio observational campaigns have shown that dSphs can provide competitive limits on the WIMP parameter space \citep{Regis14,Regis17}.  The Square Kilometer Array (SKA) and its precursors will be able to progressively close in on the scenario (b).

Statistical methods include the study of the cross-correlation of the radio synchrotron background with low-redshift DM tracers \citep{FR14} and the analysis of the 1-point probability distribution function (also called $P(D)$) of the radio synchrotron background \citep{Vernstrom15}.  The general idea behind the first approach is that the DM signals peak at redshift $<0.1$, so they can be separated from non-thermal radio emissions originating from more mundane astrophysical processes that typically trace the star formation history and peak at higher $z$.  $P(D)$ studies can probe the scenario (a) by employing data from both long and short baselines.  In this way, SKA and its precursors can indeed provide a way to distinguish between point-like and diffuse contributions to the radio synchrotron background.

\subsection{The Far-Infrared / Radio Correlation and Halos --- Eric Murphy}\label{Murphy}

Radio continuum emission is widely used as a tracer of recent star formation for galaxies both at low and high redshift, resulting from the well known tight empirical correlation between their (predominantly non-thermal) total GHz radio and far-infrared (FIR) luminosities \citep[e.g.][]{Helou85}.  Although this relation must be rooted in common dependencies on star formation, it is unclear how presumably unrelated physical processes affecting the propagation of cosmic ray  electrons and the heating of dust grains work together to yield a nearly ubiquitous correlation over many orders of magnitude in luminosity.  However, when it comes to the FIR-radio correlation, pieces of galaxies do not always behave like galaxies, and significant variations in the ratio of the emissions can be found within galaxies.   This is potentially relevant when considering origin scenarios for the radio synchrotron background.  For example, one cannot necessarily simply apply the radio-FIR correlation ratio to FIR emission to a region in the Milky Way halo and achieve the expected radio emission surface brightness from those regions.  

\begin{figure}[!ht]
\includegraphics[width=3.5in]{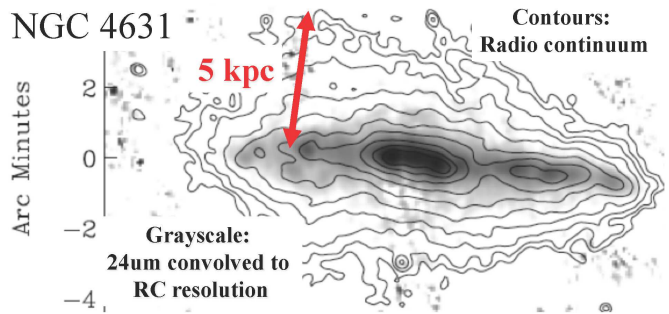}
\caption{NGC\,4631 showing 22\,cm contours overlaid onto a {\it Spitzer} 24\,$\mu$m grayscale image that has been convolved to match the 25\arcsec Gaussian radio beam; data from both images are clipped at the 3$\sigma$ level.  As is readily seen, the radio continuum emission extends much further into the galaxy halo than the warm dust emission traced by the 24\,$\mu$m intensity. }
\label{fig:n4631}
\end{figure}

One would like to investigate the properties of the interstellar medium that result in individual regions following or not following the overall radio-FIR correlation.  {\it Spitzer} revealed local correlations in the spatial distributions of 70\,$\mu$m and non-thermal radio emission in the disks of galaxies that reflect an ``age effect"; the cosmic ray electron populations of galaxies with intense star formation largely arise from recent episodes of enhanced star formation activity and have not had time to diffuse to significant distances \citep{Murphy06,Murphy08}.  The study of edge-on spiral galaxies provides a dimension to our understanding of galaxy structure and evolution unobtainable in any other way.  In particular, observationally characterizing the role of feedback processes resulting from the accretion, expulsion, and/or cycling of material between galaxy halos and disks in any significant detail is best carried out for such objects.  Gaseous halos are both the depository of galaxy feedback processes (e.g. from AGN and supernovae), and the interface between the disk's interstellar medium (ISM) and the intergalactic medium, through which infall occurs, which is required for lasting star formation in disks. H{\sc i} halo masses alone are typically a few x 10$^8$ $M_{\odot}$ \citep[e.g.][]{O07}, which, along with warm/hot ionized halos and a cycling time of order 10$^8$ years \citep{Collins02}, imply that the entire ISM of a typical spiral may be cycled through the halo in less than a Hubble time.  Of the many outstanding questions that edge-on galaxies provide a means to answer (related to the processes governing the interchange of disk/halo material), one such question is what can dust and radio continuum halos tell us about transport effects that are important for understanding the FIR-radio correlation?  

Using a combination of radio and far-infrared data for a sample of 15 disk galaxies, \citet{Murphy08} applied an image-smearing analysis to measure the propagation lengths of cosmic-ray electrons to see if the vertical diffusion into the halos of galaxies differs from radial diffusion in galaxy disks \citep[e.g.][]{Murphy06,Murphy13}.   The underlying assumption in this analysis is that the only difference between the intragalaxy morphologies of the far-infrared and radio continuum emission is due to the diffusion and energy losses of the cosmic rays because of the common origin for the cosmic ray leptons and nuclei.   To summarize the procedure, the far-infrared images are convolved by a parameterized kernel $\kappa({\bf r})$, then the residuals are computed between the free-free corrected 1.4\,GHz and smoothed infrared maps.  Due to the large range in inclination among the sample galaxies, results were compared for exponential smoothing kernels oriented in either the plane of the galaxy disk or applied isotropically.  

To determine how inclination affects the choice of smoothing kernel, the sample was separated into low ($i \leq 60\degr; N=11$) and high ($i > 60\degr; N=4$) inclination bins.  As expected, the difference for face-on galaxies appears almost negligible when using isotropic kernels or those oriented in the plane of the galaxy disks.   Conversely, if a galaxy's inclination is greater than $\sim$60$\degr$ then the orientation of the kernel seems to become important. In this case, isotropic kernels are strongly favored.  

The regular component of a galaxy's magnetic field is spread most densely throughout its {\it thin} disk; the radial diffusion of cosmic ray electrons should then preferentially occur along field lines, while vertical (out-of-plane) diffusion should require cosmic ray electrons to undergo an increased amount of cross-field diffusion. Such a scenario has been verified empirically.  Indirect estimates of diffusion coefficients for the vertical propagation of cosmic ray electrons in a galaxy's thin disk have been found to be an order of magnitude smaller than those for radial diffusion \citep{Dahlem95}.

However, in galaxies with active star formation, ordered magnetic fields can be ruptured allowing cosmic ray electrons to quickly escape their disks and form synchrotron halos \citep[e.g.][]{Hummel88}.   This decrease in vertical confinement will lead to a diffusion behavior.  A clear example of this scenario is seen in the nearly edge-on galaxy NGC\,4631 (see Figure \ref{fig:n4631}).  NGC\,4631 has a large radio halo that extends 5\,kpc beyond the vertical extent of the FIR disk \citep{Dahlem95}.  These results therefore suggest that the highly inclined sample galaxies each possess, at least to some degree, synchrotron halos in which the diffusion of cosmic ray electrons occurs on similar time-scales as those in the disk.

\subsection{Tests of Simple Galactic Origin Scenarios --- Jack Singal}\label{Singal}

This talk focused on observational constraints on a Galactic origin for the reported radio synchrotron background level.  The simplest scenario for a Galactic origin for this monopole would be an emitting, roughly spherical, halo component that exists in addition to the plane-parallel component of Galactic emission, for example as presented by \citet{SC13}.

\begin{figure}
\includegraphics[width=3.5in]{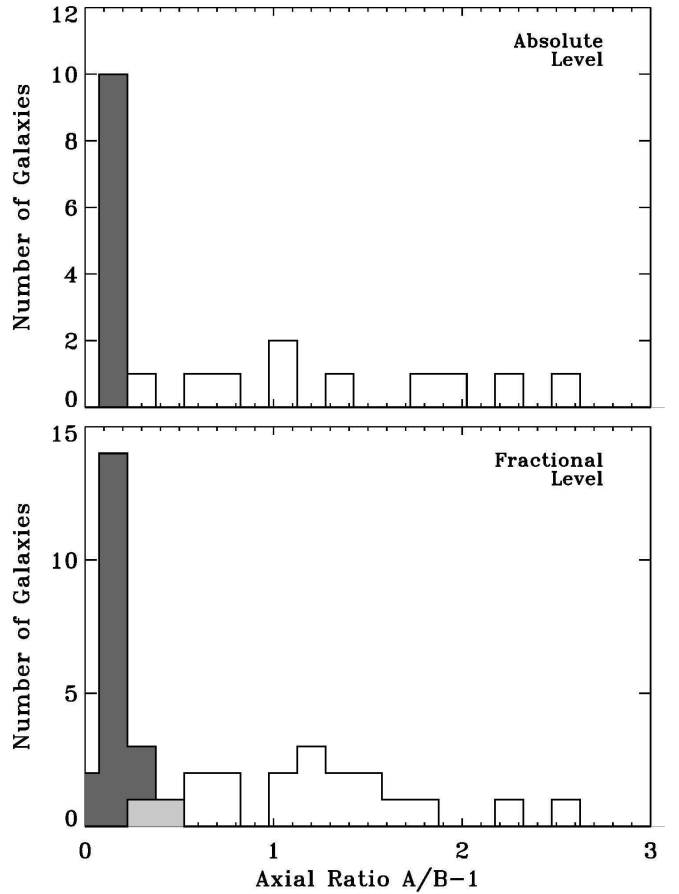}
\caption{Axial ratio of a radio emission contour in nearby edge-on spiral galaxies, capable of distinguishing between the presence or absence of a bright halo that would produce the observed level of the radio synchrotron background.  The predicted axial ratios of such theoretical halos are shown tallied in the shaded histograms, while the actual observed axial ratios are tallied in the non-shaded histograms.  These galaxies' contours are significantly more elliptical than would be the case if they had bright spherical halos capable of producing the observed level of the radio synchrotron background. From \citet{RB1} and discussed in \S \ref{Singal}.  } 
\label{SC}
\end{figure}

One test of such a model is whether emission in external spiral galaxies (presumably similar to the Milky Way) fits it.  This would be most discernable in galaxies viewed edge-on with enough resolution to differentiate a spherically shaped outer halo component from an elliptically shaped inner disk component.  \citet{RB1} examined radio emission contour maps of resolved edge-on spiral galaxies and found that all galaxies mapped with sufficient resolution were more elliptical in their outer regions than the bright halo model, indicating that if the Milky Way had a halo bright enough to account for the radio synchrotron background it would be quite anomolous.  A partial visualization of these results is shown in Figure \ref{SC}.  It is also important to note that even galaxies with extended halo emission do not generally have a bright spherical halo, as seen in e.g. Figure \ref{fig:n4631}.

Another constraint on Galactic halo emission is from the inverse-Compton emission that would accompany synchrotron emission from energetic electrons.  As a given level of synchrotron emissivity is the result of a combinaton of electron energy and magnetic field energy, the lower the magnetic field intensity the higher the electron energy must be, and vice versa, to achieve a given observed level of synchrotron.  Energetic electrons, however, also result in inverse-Compton upscattering of ambient photons to higher energies.  As quantified in \citet{Singal10}, in order to not overproduce the observed levels of the X-ray and gamma-ray backgrounds (via inverse-Compton scattering of the microwave and optical/UV backgrounds), the radio synchrotron background must originate in regions with magnetic fields of at least a few $\mu$G.  This actually under-constrains the magnetic fields if the origin were more local to the Galaxy, because the ambient level of the optical/UV energy density is higher in the vicinity of the Galaxy.  Faraday rotation measurements indicate that magnetic fields in the Galactic halo are on the order of 1\/$\mu$G \citep[e.g.][]{Taylor02}.

It is also the case that relatively simple models seem to accurately describe the large-scale structure of Galactic emission, and these models feature a halo component too dim to produce the observed level of the radio synchrotron background.  These models, discussed in \citet{Kogut11}, are (i) a cosecant of Galactic latitude fit, and (ii) the application of the observed correlation on the sky of radio and \ion{C}{2} emission.  Both fit the overall Galactic radio spatial structure well, and agree on a predicted Galactic zero-level that is much dimmer than the observed radio synchrotron background level.

\subsection{PIXIE: The Primordial Inflation Explorer and Potential RSB Constraints --- Alan Kogut}\label{Kogut2}

PIXIE (the Primordial Inflation Explorer) is a Phase A-selected proposed NASA MidEx mission for mapping CMB polarization and determining the CMB absolute spectrum \citep{PIX}.   It will consist of a Fourier transform spectrometer with polarization filters.  Emission from positions on the sky can be compared to another position, or to the emission from an absolute external calibrator modeled on the one developed for ARCADE~2 \citep{Singal11,excal}.  PIXIE will map the absolute intensity and linear polarization (Stokes $I$, $Q$, and $U$ parameters) over the full sky in 400 spectral channels spanning 2.5 decades in frequency from 30\,GHz to 6\,THz (1\,cm to 50\,$\mu$m wavelength).  In addition to being highly sensitive to the $B$-mode polarization signal and therefore a crucial probe of inflationary physics, it will be $\sim$1000 times more sensitive to CMB $y$ and chemical potential spectral distortions than the current limits from COBE FIRAS.  At this level both signals should be affirmatively detected.  These will constrain structure formation in the unvierse and energy injections into the primordial plasma at a level that will be a definitive test for the existence of warm dark matter.  PIXIE will also determine the amplitude of the cosmic infrared background to the sub-percent level.

Relevant for the radio synchrotron background, PIXIE will measure the absolute and polarization level synchrotron sky at frequencies near 30 GHz.  This will allow the determination of the fractional polarization, which will provide a definitive constraint on the local bubble origin scenario for the radio synchrotron background discussed in \S \ref{Kogut1}.

\subsection{Synchrotron Radiation as a Foreground to the Global Redshifted 21-cm Measurement by EDGES --- Raul Monsalve}\label{Monsalve}

The Experiment to Detect the Global EoR Signature (EDGES) strives to detect the cosmological 21-cm line at frequencies below 200\,MHz due to the evolution of neutral hydrogen in the IGM at redshifts $z>6$ \citep{BR10,Monsalve17}. Neutral atomic hydrogen emits the 21-cm line when the atom transitions from the upper ground state to the lower ground state. This transition corresponds to the alignment of the spins of the proton and electron going from parallel to antiparallel. Because of the redshift, the observed frequency for this spectral line emitted in the early universe is given by $\nu$=1420\,MHz / (1+$z$). For a reference redshift range $27>z>6$, the corresponding observation frequencies are $50<\nu<200$\,MHz. Therefore, the large-scale evolution of the 21-cm radiation due to different cosmological processes is expected to be imprinted as wideband features in the low-frequency radio spectrum.

Studies of the 21-cm emission focus on either the anisotropic signal, or the global absolute signal.  The 21-cm absolute spectrum is measured as a brightness temperature relative to the temperature of the background radiation, which is normally assumed to correspond to the CMB \citep[e.g.][]{Zald04}. In traditional models, before the formation of the first stars and galaxies, the 21-cm radiation emitted by neutral hydrogen in the intergalactic medium (IGM) was coupled to the CMB and, therefore, the expected differential temperature is zero. When the first sources formed, they altered the differential temperature through three main processes \citep[e.g][]{Furlanetto06}. Going from high to low redshift, these processes are the following. (1) The first sources emitted ultraviolet radiation that coupled the 21-cm radiation to the physical temperature of the IGM. At these redshifts, the IGM was colder than the CMB due to the adiabatic expansion of the universe and, therefore, this coupling takes the differential brightness temperature into absorption. (2) When a significant population of stellar remnants (neutron stars and black holes) formed, they emitted enough X-rays to heat the IGM, which in turn increased the 21-cm temperature, potentially above the CMB temperature. This leads to the end of the absorption feature and possibly to an emission feature in the spectrum. (3) As more sources appeared, their ultraviolet and X-ray radiation ionized the neutral hydrogen, gradually extinguishing the source of 21-cm radiation and reducing the differential brightness temperature to zero. Thus, the 21-cm spectrum can inform us about the time of the appearance of the first objects in the universe, as well as their characteristics and their effects on the intergalactic medium throughout ‘cosmic dawn’ and the epoch of reionization (EoR).

\begin{figure}[!ht]
\includegraphics[width=3.5in]{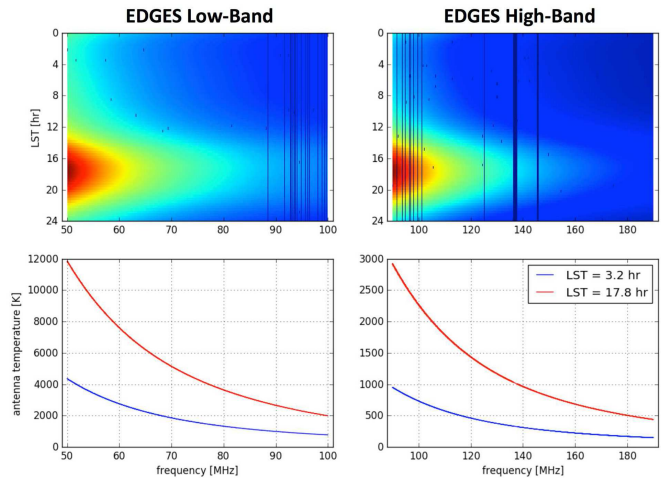}
\caption{Data from EDGES.  The top panels show the measured antenna temperature as a function of local sidereal time (LST), for a typical daily observation. In the color map used, blue, green, and red, represent low, medium, and high temperatures, respectively.  The highest temperatures are registered when the Galactic center is directly overhead, at LST = 17.8 hr.  Dark blue pixels and stripes represent data excised due to radio-frequency interference, which is most frequent in the FM radio band (87--107\,MHz) and at 137\,MHz, where ORBCOMM satellites operate.  The bottom panels show two cuts of the data sets in the top panels; specifically, they show the antenna temperature spectra at the times when most of the Galaxy is below the horizon (LST=3.2\,hr) and directly overhead (LST=17.8\,hr), respectively. }
\label{edgesfig}
\end{figure}

The brightness temperature of the sky-average, or global, 21-cm spectrum is expected to be in the range between tens and hundreds of mK. This contrasts dramatically with the intensity of the diffuse Galactic and extragalactic contributions in the MHz range which are dominated by synchrotron radiation and have temperatures between hundreds and thousands of Kelvin, as discussed throughout this report. The EDGES strategy, to try to detect and characterize the 21-cm spectrum at the mK level, consists of measuring with high accuracy the absolute brightness temperature of the sky, which represents the sum of all the radio contributions as a function of frequency \citep{RB08,Mozdzen17}. This measurement is conducted with a wide antenna beam (full-width-at half-maximum $>$70$^{\circ}$) that provides a high sensitivity to the monopole radiation component. Therefore, EDGES measurements can help to constrain the zero level of the radio contributions and, in particular, of the excess radio synchrotron background, whose existence still waits for further verification in the MHz range.

EDGES is currently on the verge of detecting the cosmological 21-cm spectrum, as a result of significant efforts to make the intrinsic passband of the instrument as smooth as possible and to improve the calibration accuracy \citep{Monsalve17b}. Calibration relies on continuously removing the instrument gain and noise offset using well characterized passive and active noise sources connected at the input of the low-noise amplifier (LNA), as well as on removing the effects of signal reflections between the antenna and LNA, and of antenna and ground losses. The estimated absolute temperature uncertainty in current EDGES measurements at frequencies above 90\,MHz is $\sim$1\,K, which represents a significant improvement over the zero-level uncertainties of the existing low frequency radio surveys discussed in e.g. \S \ref{intro} and \S \ref{Fixsen}.  Work is underway to publish updated EDGES estimates for the absolute sky temperature, and to evaluate their agreement with models for the claimed radio synchrotron background background level.  Figure \ref{edgesfig} highlights the ability of EDGES data to provide an estimate for the monopole level of the radio sky at MHz frequencies.  In addition to being of interest in its own right, a confirmed high radio synchrotron background level at these frequencies will be of direct interest to the 21-cm community. Theorists would have to incorporate into their models the possibility of a cosmological origin of this background and determine the implications for the 21-cm spectrum, as most likely the CMB would no longer be considered the only radiation background potentially present during the formation of the first generations of sources and their reionization of the IGM.

\subsection{Galactic Radio Loops --- Philipp Mertsch}\label{Mertsch}

The Galactic contribution to the radio synchrotron background is usually modeled on large, Galactic scales only. Yet, the emissivity varies on intermediate and small scales, too: a turbulent magnetic field will lead to variations of the synchrotron emissivity, mostly on scales of the coherence length of the magnetic field. In addition, this turbulence is generated by supernovae and their remnants and these could modify the emissivity on spatial scales up to hundreds of parsecs. In a sky map, such structure is conveniently characterized by the angular power spectrum.

In a first step, one can model the angular power spectrum of the Haslam 408\,MHz sky map \citep{Haslam} as the sum of a two-dimensional GALPROP \citep[e.g][]{OS13} model, the WMAP MEM free-free emission map and a contribution from unsubtracted point sources. This model exceeds the measured angular power on the largest scales (multipoles $\ell \leq 10$) by $50 \, \%$ and falls short by $50 \, \%$ on smaller scales ($10 < \ell \leq 100$). As it turns out, the small-scale turbulence in the Galactic magnetic fields only marginally contributes, since most sight lines through the synchrotron halo are much longer than the coherence length of the turbulent magnetic field.

The solution proposed in \citet{MS13} is that the $\mathcal{O}(100) \, \mathrm{pc}$ size, shells of $\mathcal{O}(10^5) \, \mathrm{yr}$ old supernova remnants compress the Galactic magnetic field and with it the cosmic ray electrons, thus shining much more brightly than the surrounding interstellar medium. Adding this component to the synchrotron model, the angular power spectrum of the Haslam map is reproduced remarkably well, to an accuracy of $10 \, \%$.  However the overall surface brightness loop component is too low to be relevant for the excess radio synchrotron background signal.

\subsection{Too Perfect?  Where are the Fluctuations? --- Gil Holder}\label{Holder}

It is known from simulations that the dark matter that dominates large scale structure in the universe is clustered.  Galaxies trace the dark matter and are therefore clustered, which has been observed at multiple wavelengths.  For example the cosmic infrared background manifests a clustering level of $\sim$10\% at scales of a few arcminutes.  If one considers mm wavelengths, the power at angular scales above $\sim$5 arcminutes ($\ell \lesssim$ 2500) is produced by the familiar CMB power spectrum, while smaller angular scales show slowly rising angular power with increasing $\ell$, resulting from the effects of emission from galaxies \citep{George15}.  The latter is a combination of shot noise in galaxy number and the intrinsic clustering from large scale structure.  It is natural to then ask what observational constraints exist at these scales at lower frequencies where the radio synchrotron background is consequential.  

\begin{figure}[!ht]
\includegraphics[width=3.5in]{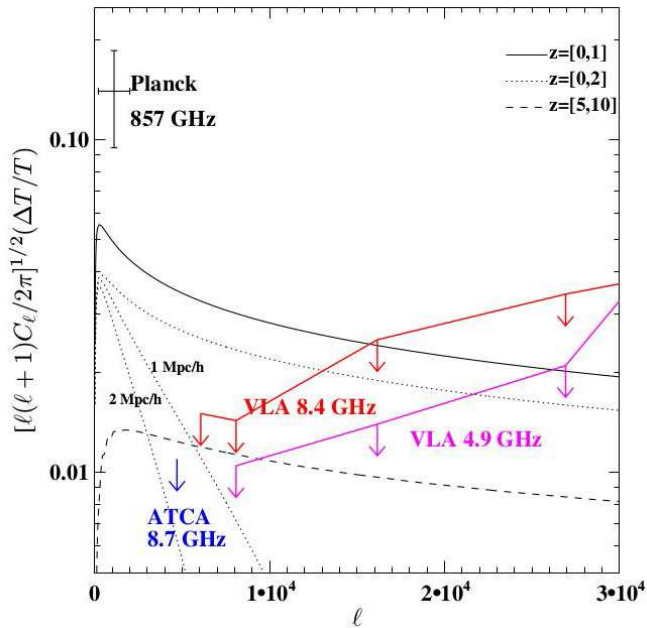}
\caption{Expected angular power from clustering for several ranges in redshift for the contributions to the unresolved radio background, as well as the observed upper limits on angular power, as discussed in \S \ref{Holder}.  For reference, the
amplitude inferred for the cosmic infrared background measured by Planck
is also shown. For the redshift interval $z$=0--2 (dotted), the effect of each source being extended is shown: top to bottom are FWHM$_{\rm smooth}$ = 0, 1, 2$h^{−1}$ comoving Mpc.  Reproduced from \citet{Holder14}.}
\label{anis}
\end{figure}

As presented in \citet{Holder14}, observations with the VLA at 8.4\,GHz \citep{Partridge97} and 4.9\,GHz \citep{Fomalont88}, and the Australia Compact Telescope Array at 8.7\,GHz \citep{Sub00}, originally carried out as searches for CMB anisotropies, have placed upper limits on the angular power at arcminute scales at these frequencies.   \citet{Holder14} calculated the expected angular power resulting from the radio synchrotron background level reported by \citet{Fixsen11} if the sources were clustered according to the observed matter power spectrum in the universe, for the responsible sources having various distributions of redshifts.  That work showed that the angular power would exceed the observed limits for the sources being of galaxy scale and located at redshifts less than $z \sim 5$, as illustrated in Figure \ref{anis}.

This constraint suggests that the sources of the radio synchrotron background, if it is at the level reported by e.g. \citet{Fixsen11}, must either be extended so as not to exhibit clustering power on arcminute scales (corresponding to physical sizes of a few Mpc), or be predominantly at high ($z\gtrsim$5) redshifts.

\subsection{Attempting to Measure the Power Spectrum of Radio Anisotropies by the Quadratic Estimator --- Lienong Xu}

The angular power spectrum technique is a standard tool for understanding the formation of large-scale structure. For example substantial cosmological information has been successfully revealed by the CMB and cosmic infrared background (CIB) angular power spectra. Motivated by the well known FIR-radio correlation of star-forming galaxies, the power spectrum in radio frequencies should contain large-scale structure content as well. Moreover, the radio power spectrum could also probe structure like the synchrotron cosmic web, which could account for the missing baryons in the observable universe \citep{Vernstrom17}.  If the contribution to the radio synchrotron background is dominated by discrete extragalactic point sources like the CIB, the clustering of the sources, i.e. the amplitude of radio power spectrum, should be at about the ten percent level \citep{Holder14}. Therefore careful estimation of the radio power spectrum is an important aspect of understanding the RSB. 

One of the factors which make estimation of the radio power spectrum difficult is the reliability of the radio images, since radio images rely on the sampling of the interferometric visibility measurement in Fourier space. However the power spectrum is naturally defined in Fourier space. Hence in principle, the radio power spectrum can be directly estimated from interferometric data without the intermediate stage of making maps in real space.  In this work, a quadratic estimator is implemented for the radio power spectrum in order to achieve unbiased, minimum variance parameter estimation. The project is currently at the simulation stage, along with a basic check with Australia Compact Telescope Array (ACTA) data from which a realistic point source power spectrum has been calculated. The noise level estimation in this procedure is a more difficult aspect than with the usual cross correlation technique.  Various null tests remain to be investigated extensively before claiming the complete verification of this framework.  Nevertheless, the hope is to be able to measure the clustering in existing radio data, thereby addressing the question of the level of clustering in the radio synchrotron background.

\section{Discussion}\label{conc}

One consensus of the workshop, as brought out in the discussion and question sessions, is that there is no tension between the ARCADE~2 measurements and the several low frequency degree-scale resolution radio maps with an absolute zero level, but profound tension between these results and extragalactic source counts from deep interferometric measurements on one hand and a suite of (but by no means all) Galactic diffuse synchrotron models and constraints on the other.  There are essentially three possible scenarios at this point.

(i) A new, absolutely calibrated zero level measurement shows a lower radio synchrotron monopole level, and that the levels measured by ARCADE~2, because of limited sky coverage, and the low frequency radio maps, because absolute zero-level  calibration was not a primary goal, were in error.  In this case there would be no ``excess'' signal above that which would be expected from the simple Galactic models discussed in e.g. \S \ref{Singal} and the extragalactic source counts discussed in e.g. \S \ref{Condon}.

(ii) The level of the radio synchrotron background as determined from ARCADE~2 and low frequency sky maps as reported by e.g. \citet{Fixsen11} is correct and the excess signal is primarily Galactic, which would fundamentally challenge our understanding of many aspects of the interstellar medium of the Milky Way, and the Milky Way's very status as a typical spiral galaxy.

(iii) The level of the radio synchrotron background as determined from ARCADE~2 and low frequency sky maps as reported by e.g. \citet{Fixsen11} is correct and the excess signal is extragalactic, which would make it the most interesting photon background in the sky at the moment, because every other photon background currently: (a) agrees with source counts; (b) has been resolved into their respective sources; and (c) their anisotropies correlate with the large scale structure of optical sources in the universe, whereas the radio synchrotron background would be anomalous in all three of those aspects.

There are reasons to consider all three possibilities (i)-(iii) at present, although the chances of the first one seem to be narrowing slightly in light of EDGES data.  The spectrum of the radio synchrotron background essentially matching that of the average of extragalactic sources is a point in favor of the third possibility, while the ability of GALPROP modeling to produce the observed level (albeit with a large emitting halo, as discussed in \S \ref{Orlando}) is a point in favor of the second possibility.

The workshop reached a strong consensus on the importance of having at least one new, purpose-built, absolutely calibrated zero level measurement to confirm (or dispute) the level of the radio synchrotron background obtained from ARCADE~2 and the low frequency radio surveys.  The options for such a measurement can be broadly classified into the following categories:

(1) A measurement at around 120\,MHz, where the radiometric temperature of the radio synchrotron background is equal to that of the ground, greatly reducing the measurement complexity surrounding ground pick-up.  Such a measurement could have a simple, chicken-wire ground plane.

(2) A measurement at around 300\,MHz, consisting of a purpose-built receiver mounted on the Green Bank Telescope, which features an unblocked aperture.  At this frequency the effect of tropospheric water vapor fluctuations is not important and the location in the National Radio Quiet Zone provides protection against radio frequency interference.  Such a setup would only take a few days to map the entire accessible portion of the sky.

(3) A measurement in the GHz range with greater sky coverage than ARCADE~2, and possibly at a frequency high enough to determine whether the radio synchrotron background spectrum rolls off above 10\,GHz, as the expected roll-off is different for Galactic and extragalactic sources.

While participants agreed that all of these measurements are potentially worthwhile, option (2) was thought to be the one that could be accomplished in the shortest time frame.

Another point raised is that any new speculation going forward for the origin of the radio synchrotron background should be thought of in tandem with a way to test such a hypothesis with either interferometry at some angular scale or an absolutely calibrated zero-level measurement at some frequency.  

The workshop was unanimously perceived as a success and worthwhile endeavor by the participants, and, along with the consensus on the need for an additional measurement, there was agreement on the importance of the topic and the desirability of a future dedicated meeting to discuss progress.\\

\acknowledgments
{\bf Acknowledgments}\\
We acknowledge support from National Science Foundation award number 1557628, and from the University of Richmond College of Arts and Sciences.  The National Radio Astronomy Observatory is operated by Associated Universities Inc. under Cooperative Agreement with the National Science Foundation.

\end{document}